**Nanostructured Germanium with >99 % Absorption at 300-1600 nm Wavelengths**

*Toni P. Pasanen\*, Joonas Isometsä, Moises Garin, Kexun Chen, Ville Vähänissi, and Hele Savin*

Dr. Toni P. Pasanen, Joonas Isometsä, Dr. Kexun Chen, Dr. Ville Vähänissi, Prof. Hele Savin
Aalto University, Department of Electronics and Nanoengineering, Tietotie 3, 02150 Espoo, Finland
E-mail: toni.pasanen@aalto.fi

Dr. Moises Garin
Aalto University, Department of Electronics and Nanoengineering, Tietotie 3, 02150 Espoo, Finland
Universitat de Vic – Universitat Central de Catalunya, Department of Engineering, c/ de la Laura 13, 08500 Vic, Spain
Universitat Politècnica de Catalunya, Gran Capità s/n, 08034 Barcelona, Spain

**Abstract**

Near-infrared (NIR) sensors find numerous applications within various industry fields, including optical communications and medical diagnostics. However, the state-of-the-art NIR sensors made of germanium (Ge) suffer from rather poor response, largely due to high reflection from the illuminated device surface. This work demonstrates a method to increase the sensitivity of Ge sensors by implementing nanostructures to the wafer surfaces. The absorbance of nanostructured Ge wafers is measured to be >99 % in the whole UV-VIS-NIR spectrum up to 1600 nm wavelength, which is a significant improvement to bare Ge wafers that reach absorption of only 63 % in maximum. The process is shown to be capable of producing uniform nanostructures covering full 100-mm-diameter substrates as well as wafers with etch mask openings of different sizes and shapes, which demonstrates its applicability to CMOS sensor manufacturing. The results imply that nanostructured Ge has potential to revolutionize the sensitivity of Ge-based sensors.

**Main text**

Near-infrared (NIR, wavelength 750–3000 nm) sensors are a subject of wide interest within various industry fields, including optical communications,[1] medical diagnostics,[2] night-vision,[3] and spectroscopy[4] just to name a few. While silicon-based devices are the mainstream for ultraviolet (UV) and visible-light detection, their applicability is limited to wavelengths shorter than ~1000 nm by the band gap of silicon.[5] Consequently, for NIR wavelength detection, silicon can be replaced by a smaller band gap semiconductor, such as germanium (Ge), which can absorb radiation until ~1600 nm.[5] The selection of Ge would be favorable by its compatibility with CMOS manufacturing lines and lower price compared to its alternatives, such as indium-gallium-arsenide (InGaAs).[6]

It is well known, however, that reflection is one of the main factors that limits the response of Ge sensors.[7] Indeed, such devices conventionally have highly-reflective bare surfaces or only a single-layer anti-reflection (AR) coating optimized for NIR, which has very limited performance for other wavelengths.[8] Moreover, texturing of Ge surfaces has remained challenging due to the lack of standard techniques, such as random pyramids[9] or acidic texturing[10] that are commonly used for silicon, for that purpose.[11] The sensitivity of such devices can hence be greatly improved by enhancing the absorption of radiation. This could potentially be realized by generating a gradually changing refractive index from air to the substrate on the illuminated device surface. Instead of using multi-layer AR coatings, an interesting alternative is to use structures that have dimensions similar to the wavelength of the detectable radiation.[12] Such nanostructures have recently been applied to silicon-based photosensors, resulting in record-high sensitivity to UV and visible radiation.[13] Application of similar nanostructures to germanium sensors could extend the superior sensitivity to the NIR region, and indeed, promising preliminary results on NIR sensors with a textured surface have recently been published.[14–16]

As learnt from research on silicon nanostructures, which are often called black silicon (b-Si) due to the negligibly small reflectance in the visible range, the nanostructure morphology is critical for device performance. Indeed, nanostructures with porous or rough sidewalls[17] or with high aspect ratio[18] have resulted in excessive surface recombination despite surface passivation, whereas excellent performance has been achieved using nanostructures with smooth sidewalls and only modestly increased surface area.[19,20] Germanium nanostructures have earlier been realized by nickel-catalyzed synthesis,[21] electrochemical etching,[22] and Bosch process-based dry etching,[23] but the morphology of the surface structures has been non-optimized for optoelectronic applications. Additionally, previous research has been performed only on small substrate pieces, which is impractical for the manufacturing of actual devices, especially since the surface area that is exposed to the etching chemistry may affect the process.[24]

The morphology and properties of nanostructured surfaces can be adjusted more accurately by reactive ion etching (RIE).[25] A preliminary study on RIE-fabricated Ge nanostructures has focused on a chlorine ($Cl_2$)-based process.[26] However, $Cl_2$ sets demanding requirements for the process equipment and gas handling due to its high toxicity.[27] Sulfur hexafluoride ($SF_6$)-based processes, which are commonly used for etching of silicon,[28] have earlier been considered inapplicable for the fabrication of Ge nanostructures.[26] Moreover, the process developed in [26] relied on capacitively-coupled plasma (CCP) alone. Such process always needs to balance between minimized reflectance and ion bombardment-induced surface damage.[20] An inductively-coupled plasma (ICP) configuration would avoid the plasma-induced damage entirely.[29]

In this work, we demonstrate a reproducible $SF_6$-based ICP-RIE process for the fabrication of Ge nanostructures for NIR sensors. The process performance is tested over the whole 100-mm-diameter substrate area as well as on wafers with etch mask openings of different sizes and shapes to mimic nanotexturing of actual sensors. We investigate the morphology of the nanostructures and study the uniformity of the nanostructure dimensions over the wafer area. Finally, the optical properties of nanostructured Ge are characterized by reflectance and absorbance measurements. The optical performance is compared to bare Ge wafers and b-Si to evaluate the potential of nanostructured Ge for boosting the detector performance.

The experiments were performed on gallium-doped (p-type) 100-mm-diameter Ge wafers, which were 175 µm thick and had a resistivity of 0.01-0.05 Ωcm. The polished wafer surface with (100) orientation was exposed to a $SF_6$-based ICP-RIE process (Oxford Instruments Plasmalab System100) at -120 °C for 15 min. The gas flows for $SF_6$ and oxygen ($O_2$) were 25 and 33 sccm, respectively, while the chamber pressure was kept at 10 mTorr. The power of the ICP source was 1000 W, whereas the CCP power was set to as low as 4 W to avoid ion bombardment-induced damage. Another similar wafer was coated with a 22 nm thick atomic-layer-deposited (ALD) aluminum oxide ($Al_2O_3$) etch mask, which was patterned by photolithography and phosphoric acid-based commercial aluminum etchant (Honeywell PWS 80-16-4 (65)). Subsequently, nanostructures were fabricated on the opened areas using the same ICP-RIE process to mimic nanostructuring of sensor active areas.

The nanostructures were characterized by scanning electron microscopy (SEM, Zeiss Supra 40) and atomic force microscopy (NT-MDT NTEGRA) from different locations on the wafer. The amount of Ge consumed by the process in vertical direction was determined by profilometry (Dektak XT) as the step height of the opened areas on the patterned wafer. Finally, the optical properties of the nanostructured surfaces were evaluated by integrating sphere-based reflectance ($R$) and transmittance ($T$) measurements (Agilent Cary 5000). The absorptance of the wafers was derived from $1 - R - T$.

The developed nanostructuring process was first applied to bare Ge wafers. **Figure 1a** presents a photograph of a nanostructured 100-mm-diameter wafer, which appears uniformly black to the naked eye. The deep black appearance over the whole wafer surface highlights the heavily reduced reflectance, at least for visible light, and demonstrates that the process is well suited for large substrate areas. The surface also retains its darkness when the wafer is tilted, as determined by the eye, which agrees with earlier reports on silicon nanostructures.[13,30] This property significantly enlarges the acceptance angle of radiation sensors compared to devices with a bare front surface or an AR coating.[13,30] The heavily reduced reflectance of nanostructured surfaces can be explained by the minuscule size of the nanostructures: they provide a gradual increase in refractive index from that of air to that of germanium, and hence, virtually eliminate the optical interface.[12]

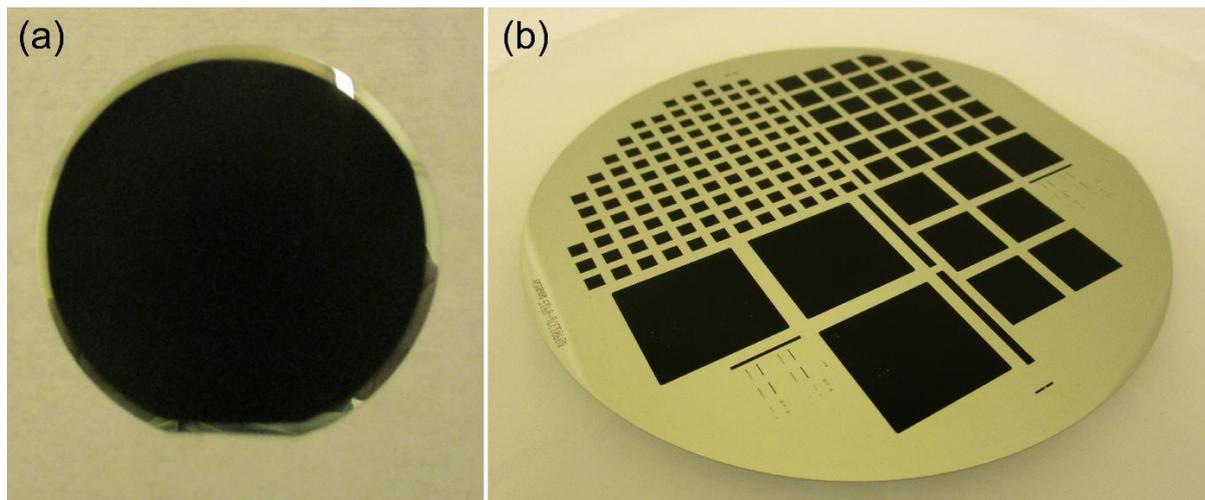

**Figure 1.** Photographs of 100-mm-diameter Ge wafers with (a) nanostructures on the whole surface, showing a uniformly black appearance, and (b) nanostructures fabricated on smaller areas with various sizes and shapes using an $Al_2O_3$ etch mask, which mimics nanostructuring of actual devices.

Since actual devices typically allow texturing only on the active areas, and not below contacts, the nanostructuring process was applied also on Ge wafers protected by an etch mask with openings of various sizes and shapes. **Figure 1b** shows such wafer with several types of rectangle-shaped nanostructured areas, which all appear uniformly dark independent of their location on the wafer. The results demonstrate that the process is applicable for different areas exposed to the etching chemistry, which is not self-evident due to the loading effect in plasma etching,[24] but is a prerequisite for sensors manufacturing.

To ensure efficient graded index effect also for NIR wavelengths in addition to UV and visible light, the nanostructures were designed to have slightly larger dimensions than the structures used in the state-of-the-art b-Si photosensors.[13] **Figure 2** shows an AFM image of the nanostructures scanned over a 10 x 10 $\mu m^2$ area. The AFM data is smoothened to reduce the effect of noise, which results in slight rounding of the valleys between the nanostructures. The figure reveals that the surface structure consists of germanium spikes with varying shapes and sizes in the nm-scale. Indeed, it is characteristic to the ICP-RIE process to produce randomly distributed nanostructures, the dimensions being statistically determined by the applied process parameters.[25] The height and width distribution can be more accurately characterized from cross-sectional SEM images, one of which is shown as an inset in Figure 2. With the applied process, the typical height of the nanostructures varies from 400 to 900 nm and their width is between 100 and 350 nm.

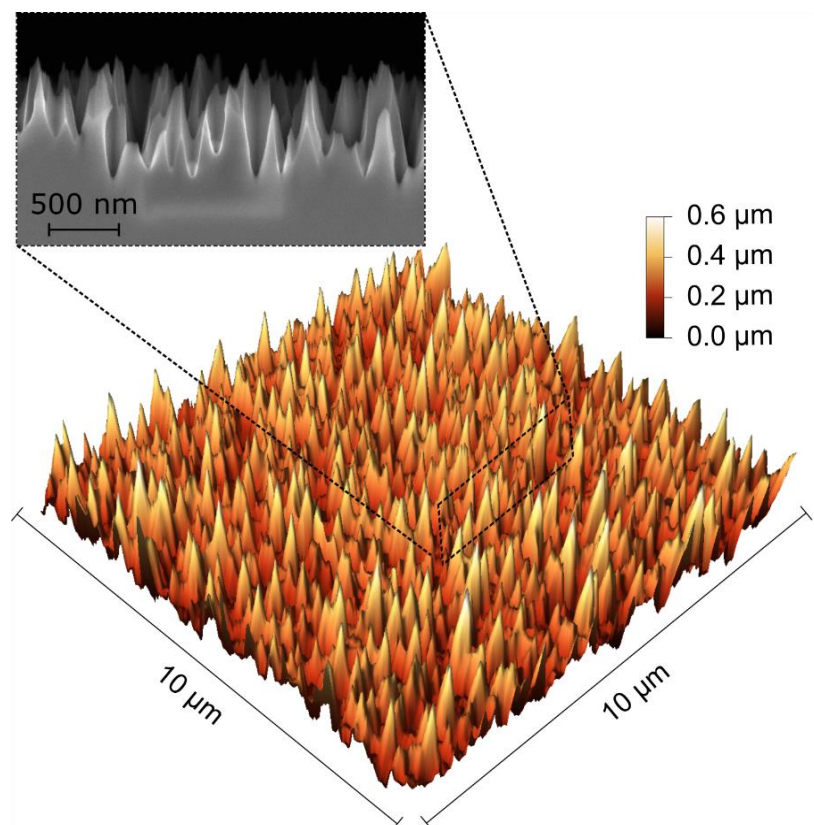

**Figure 2.** Atomic force microscopy (AFM) image of the nanostructured Ge surface. The inset shows a cross-sectional scanning electron microscope (SEM) image of the nanostructures.

The size and shape distribution of the nanostructures is controlled by the process parameters, including the $SF_6/O_2$ gas flow ratio, ICP and CCP power, chamber pressure, and etching time.[25] The process is driven by chemical etching: fluorine radicals react with germanium, forming volatile $GeF_4$, whereas oxygen contributes to the formation of a rapidly-desorbing $GeO_xF_y$ layer.[31] With high $SF_6$ concentration in the gas mixture, germanium is etched isotropically, preventing the formation of nanostructures. On the other hand, nanostructure formation is suppressed also when the $O_2$ flow is overly high with respect to $SF_6$, since the etch rate of germanium is heavily reduced.[32] Hence, the optimization of the gas composition is critical. In this work, a $SF_6$ to $O_2$ gas flow ratio of 25 to 33 sccm was found to be optimal for producing uniform nanostructures throughout the wafer. Similar to $SF_6$ flow, the ICP power affects the density of fluorine radicals, and consequently, increased ICP power results in accelerated etching of germanium.[33]

The nanostructure morphology is affected also by the energy and directionality of the ions that bombard the substrate, which are determined by the CCP power and chamber pressure. Increased CCP power and/or decreased pressure promote more directional etching, which results in dense arrays of sharp needle-shaped nanostructures with close to vertical sidewalls.[25] However, too directional ion bombardment results in anisotropic etching of the substrate without nanostructure formation. Indeed, in our experiments the process described above with CCP power increased from 4 W to 6 W resulted in a bright area in the middle of the wafer, indicating the lack of nanostructures. The depth of the nanostructures is affected by several process parameters, but it can be controlled most straightforwardly by the etching time. In our experiments, a five-minute shorter etch process resulted in smaller nanostructures and brighter appearance of the wafer.

The SEM images indicate that the morphology of the nanostructures could be well suitable for high-performance devices, where minimization of surface recombination is essential. More specifically, the needle-like structures have smooth sidewalls, which minimizes the surface area, and consequently, recombination, while they yet efficiently reduce reflectance. Furthermore, the needles have moderate aspect ratio, i.e., height compared to their width, which suggests that they can be coated conformally by ALD thin films. Indeed, the nanotexture morphology is very similar to b-Si nanotexture obtained by similar methods, where excellent surface passivation has been achieved by ALD $Al_2O_3$,[19] leading to superior b-Si solar cells[30] and photodiodes.[13]

In addition to providing accurate control over the morphology, it is important that the nanostructuring process does not consume too large amount of Ge, which might affect the fabrication process and the operation of the sensor and would require the use of thicker wafers. The nanostructuring process removed only ~4.4 µm of germanium in vertical direction, as determined by profilometry from the center of the wafer, which equals to an etch rate of ~300 nm/min. The result was also confirmed by SEM. The process is thus capable of producing sub-micron structures without significantly thinning the substrate and, hence, it is applicable also to thin wafers and Ge films with at least a few µm thickness, which is important for reducing material costs. Another etch profile scan revealed that the etch depth is slightly smaller (~3.9 µm) at a 5 mm distance from the wafer edge. Also SEM images showed that although the nanostructures remain very homogeneous at least until ~10 mm from the wafer edge, they are somewhat shorter near the area that was under the clamp during the etching process. Nevertheless, the slight variation in nanostructure size or etch depth has no effect on the dark appearance of the wafer surface as determined by the naked eye.

The optical performance of the nanostructured surfaces is evaluated quantitatively in **Figure 3**. The reflectance of the nanostructured Ge wafer is only 0.85 % on average in the wavelength range from 250 to 1800 nm, while that of the flat surface is more than 40 %. Figure 3a confirms that, in addition to the visible range, the reflectance remains negligible also for NIR wavelengths, which cannot be seen by the eye. Hence, the behavior of reflectance as a function of wavelength is significantly different from that of AR-coated surfaces, which can be optimized only for a single wavelength.[8] The graded index effect impacts on a wide range of wavelengths from UV to NIR due to the wide distribution of nanostructure dimensions on the wafer surface (Figure 2). The local maxima in the reflectance of the polished sample at ~260 and 600 nm wavelengths are characteristic to single-crystalline bulk germanium.[34]

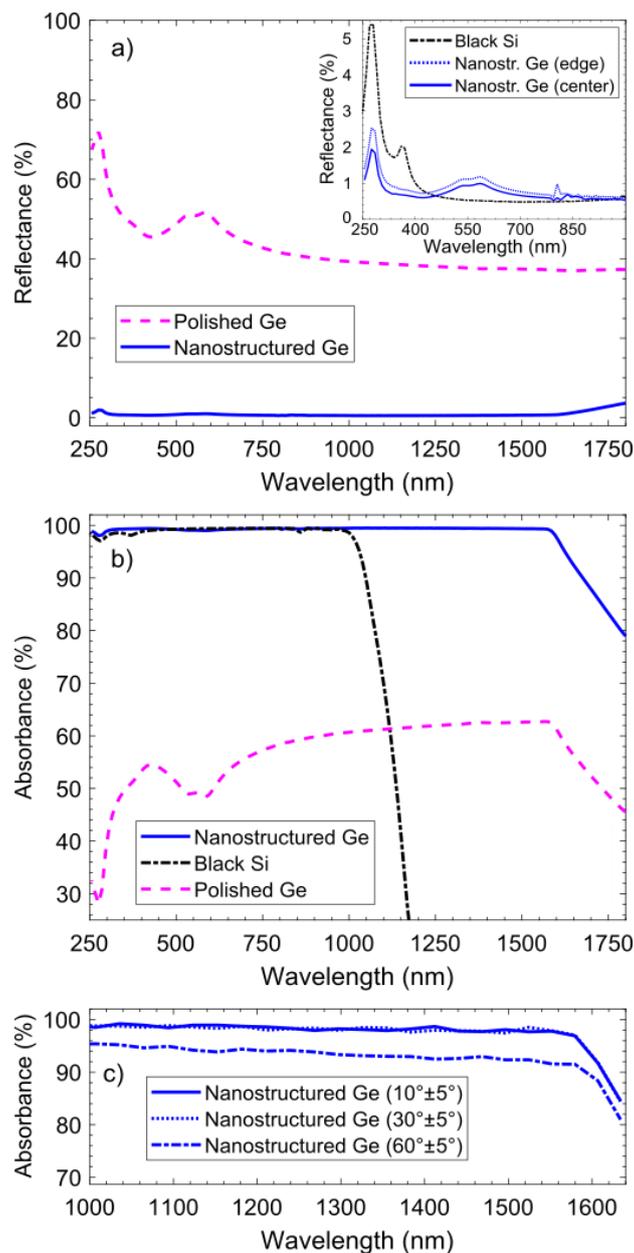

**Figure 3.** (a) Reflectance of nanostructured and polished Ge wafers in the wavelength range from 250 nm to 1800 nm. The inset compares the reflectance of nanostructured Ge to b-Si fabricated with the same method (data from [35]). It also demonstrates the uniformity of the optical properties: the spectra have been measured from the center and near the edge of the 100-mm-diameter nanostructured Ge wafer. (b) Absorbance of nanostructured and polished Ge wafers compared to b-Si (data from [13]). (c) Absorbance of nanostructured Ge wafers at different incidence angles. Note the different scale of the x-axis in figure (c).

Although the nanostructures were observed to be slightly smaller close to the wafer edge by SEM, the inset in Figure 3a shows that the difference in reflectance between the center and the edge of the 100-mm-diameter wafer is negligible. This demonstrates that the process is capable of producing nanostructures with uniform properties irrespective of the position on the wafer, and hence, is applicable to manufacturing of nanostructured Ge sensors on large wafer areas.

Another relevant optical parameter, even more than reflectance, for the sensor performance is the absorbance of radiation, which directly affects the device sensitivity. The extremely low reflectance of the nanostructured wafers results in an absorbance higher than 99 % in the UV, visible, and NIR regions until 1600 nm wavelength (Figure 3b), after which the absorption coefficient of Ge starts to reduce.[5] The improvement compared to polished Ge or b-Si wafers is evident: the former reaches an absorption of only 63 % at the maximum, while the latter is virtually transparent for wavelengths larger than 1000 nm due to the smaller band gap of silicon.[5] Furthermore, Figure 3c verifies that the absorbance remains excellent also at large incidence angles. More specifically, the absorbance is not affected by a 30° tilt and it reduces only by a few percent absolute at an incidence angle as large as 60°. The increased absorbance should be visible in the sensor performance as improved external quantum efficiency and response, similar to reported before for b-Si photosensors[13].

This work demonstrated a reproducible ICP-RIE process to fabricate Ge nanostructures for the application in photosensors. The process was shown to be capable of producing uniform nanostructures on both large-area substrates and patterned wafers, which demonstrates its high applicability to CMOS sensor manufacturing. The morphology of the Ge nanostructures was designed for maximized absorption, consequently, the absorbance of nanostructured Ge wafers was >99 % in the 300-1600 nm wavelength range and remained high also at large incidence angles. This result confirms that the basic requirement for achieving photosensors with nearly ideal response from UV to NIR is fulfilled, i.e., all incident photons enter the substrate generating a high amount of charge carriers. The other important requirement is obviously the efficient collection of the generated carriers, which heavily depends on surface passivation. Indeed, the surface recombination at the Ge nanostructures should be addressed next. The morphology of the nanostructures presented here suggest that ALD could be the key technology to achieve excellent passivation of nanostructured Ge surfaces, similar to what has been reported earlier for silicon.

**Conflict of Interest**

The authors declare no conflicts of interest.

**Acknowledgements**
The authors acknowledge the provision of facilities by Aalto University at OtaNano – Micronova Nanofabrication Centre. Nicklas Anttu and Caterina Soldano are acknowledged for the help in angle-dependent absorbance measurements and AFM characterization, respectively. The work was funded through the ATTRACT project funded by the EC under Grant Agreement 777222 and by Business Finland through project RaPtor (687/31/2019). The work is related to the Flagship on Photonics Research and Innovation "PREIN" funded by Academy of Finland.

**Table of contents entry**

Absorbance of germanium substrates is increased to >99 % in the whole UV-VIS-NIR spectrum up to 1600 nm wavelength by nanostructuring the surfaces. The developed nanostructure fabrication process is demonstrated to be highly applicable to CMOS sensor manufacturing. The results imply that nanostructured germanium has potential to revolutionize the sensitivity of Ge-based sensors.

**Keyword:** nanostructures

Toni P. Pasanen*, Joonas Isometsä, Moises Garin, Kexun Chen, Ville Vähänissi, and Hele Savin

**Nanostructured Germanium with >99 % Absorption at 300-1600 nm Wavelengths**

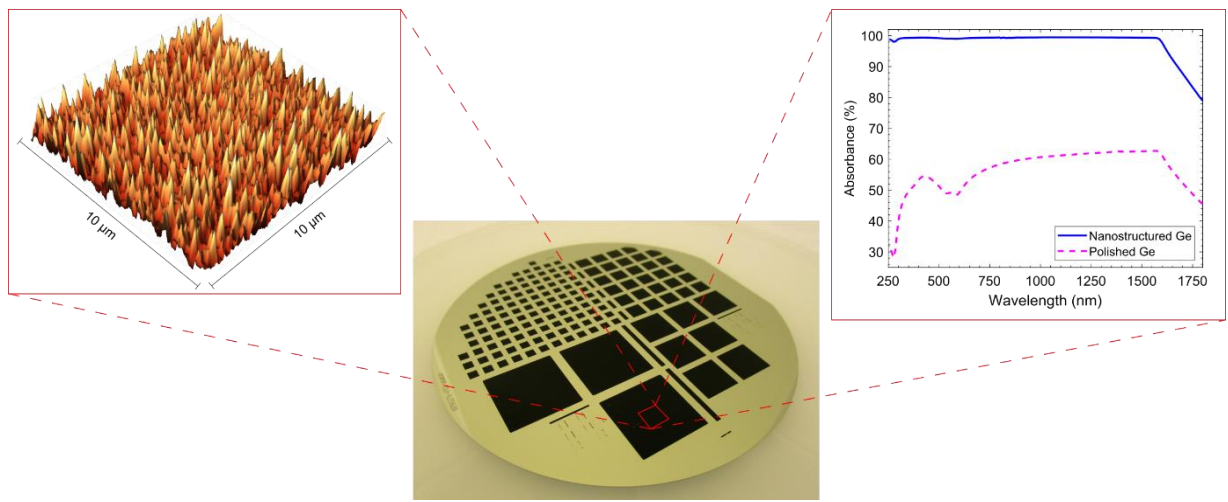